# Imaging the field inside nanophotonic devices

*Tal Fishman[1†], Urs Haeusler[2,3†], Raphael Dahan[1], Michael Yannai[1], Yuval Adiv[1], Tom Lenkiewicz Abudi[1], Roy Shiloh[2], Ori Eyal[1], Peyman Yousefi[2], Gadi Eisenstein[1], Peter Hommelhoff[2], and Ido Kaminer[1]*

*† equal contributors*
*1 Department of Electrical and Computer Engineering, Technion – Israel Institute of Technology, Haifa 32000, Israel*
*2 Department of Physics, Friedrich-Alexander-Universität Erlangen-Nürnberg (FAU), Staudtstraße 1, Erlangen 91058, Germany*
*3 Cavendish Laboratory, University of Cambridge, JJ Thomson Avenue, Cambridge CB3 0HE, UK*

*Author e-mail address: ftal@technion.ac.il, kaminer@technion.ac.il*

## Abstract

Controlling optical fields on the subwavelength scale is at the core of any nanophotonic device. Of particular interest are nanophotonic particle accelerators that promise a compact alternative to conventional radiofrequency-based accelerators. Efficient electron acceleration in such compact devices critically depends on achieving nanometer control of electron trajectories by precisely designed optical nearfields inside the device. However, these nearfields have so far been inaccessible due to the complexity of the devices and their geometrical constraints, hampering efforts to design and optimize future nanophotonic particle accelerators. Here we present the first measurement of the field distribution inside a nanophotonic accelerator. We develop a novel microscopy approach based on photon-induced nearfield electron microscopy (PINEM) to achieve frequency-tunable deep-subwavelength imaging of the nearfield inside nanophotonic accelerators. We compare the two leading designs of nanophotonic accelerators, also known as dielectric laser accelerators (DLAs): a dual-pillar structure with distributed Bragg reflector and an inverse-designed resonant structure. Our experiments are complemented by full 3D simulations, unveiling surprising deviations from the expected designs, showing complex field distributions related to intricate 3D features in the device and its fabrication tolerances. We further envision a tomography method to image the 3D field distribution, key for the future development of high-precision and hence high-efficiency DLA devices as well as other nanophotonic devices.

# Introduction

Particle accelerators are critical components in modern industrial, academic and medical infrastructure. Their applications range from X-ray generation for cancer treatment and medical imaging to the high-end experiments of high energy physics[1-3]. Current accelerators are based on radiofrequency technology that makes them large and expensive. Shrinking the size of accelerators could open new possibilities for numerous applications, including electron diffraction and microscopy, portable medical X-ray sources, and coherent probes for quantum information science[4,5].

One approach toward compact particle accelerators facilitates efficient interaction between the particles and light by using short laser pulses and nanophotonic devices built from dielectric materials[6]. An important advantage of these nanophotonic dielectric laser acceleration (DLA) devices, compared to metal or superconducting radiofrequency accelerators, lies in the two order of magnitude higher damage threshold of dielectrics at optical frequencies[7]. Together with their small footprint on the order of millimeters and the use of off-the-shelf infrared laser sources as the energy source for acceleration, DLA promises an economical and compact alternative to radiofrequency accelerators. The operation of DLA devices is based on the inverse Smith-Purcell effect[8-10], a process of stimulated absorption and emission of photons by an electron that passes close to a periodic structure. Over the last sixty years, various methods for enhancing the fundamental electron-light interaction in DLA structures have been proposed, see for example[6,11-18]. The first successful DLA demonstrations from 2013 showed a few dozens of MeV/m acceleration gradient for sub-relativistic electrons[8,19] and up to 250 MeV/m for relativistic electrons[20]. Since then, continuous improvements in DLA designs and laser coupling efficiency, benefitting from the mature technology of silicon photonics, led to the demonstration of acceleration gradients up to and above the GeV/m scale[17,21-27]. The current bottleneck for pushing forward DLA technology is in the precise control of the electron trajectories, which depend on the exact nearfield distribution inside the DLA. This high degree of control is necessary for longer and more modular acceleration structures and for the concatenation of multiple such devices.

Further optimization and advanced electron beam focusing techniques, such as alternating phase focusing[18,28,29], require a detailed understanding of the 3D nearfield distribution inside the accelerator structure. While simulations can provide extensive insight into the field distribution, they depend on the precise 3D knowledge of the fabricated structure dimensions and laser beam parameters, which are rarely fully accessible. Consequently, direct characterization of the field distribution is of paramount importance. However, until now, no experimental method has been able to measure the actual field distribution inside nanophotonic accelerators.

In this work, we demonstrate the deep-subwavelength characterization of the accelerating electric field inside a DLA channel. Using a transmission electron microscope (TEM), the electrons traverse the DLA channel and image the transverse distribution of the electric field inside it (Fig. 1). We also extract the spectral response of the DLA by scanning the wavelength of a continuous wave (CW) laser excitation, thus **acquiring a deep-subwavelength nearfield image at each wavelength**. The field information is extracted from the electron energy shift at each point by filtering over the electron energy, distinguishing the absorption and emission of individual photon quanta via energy-filtered transmission electron microscopy (EFTEM).

We compare the two leading designs of silicon-based DLA structures: (1) the established design based on two rows of pillars[17] with a distributed Bragg reflector on one side (Fig. 1b)[26,30], and (2) an inverse-designed structure with an enclosed acceleration channel for resonant enhancement (Fig. 1c)[27,31,32]. We explore the sensitivity of both devices to small changes in the optical wavelength. We find that the actual field distribution differs from the design target and accordingly affects the optimal performance of the DLA structures. Comparing the experimental results to the theoretical predictions brings new insights into the current DLA designs, which is crucial for the design of future, more ambitious and complex DLA structures.

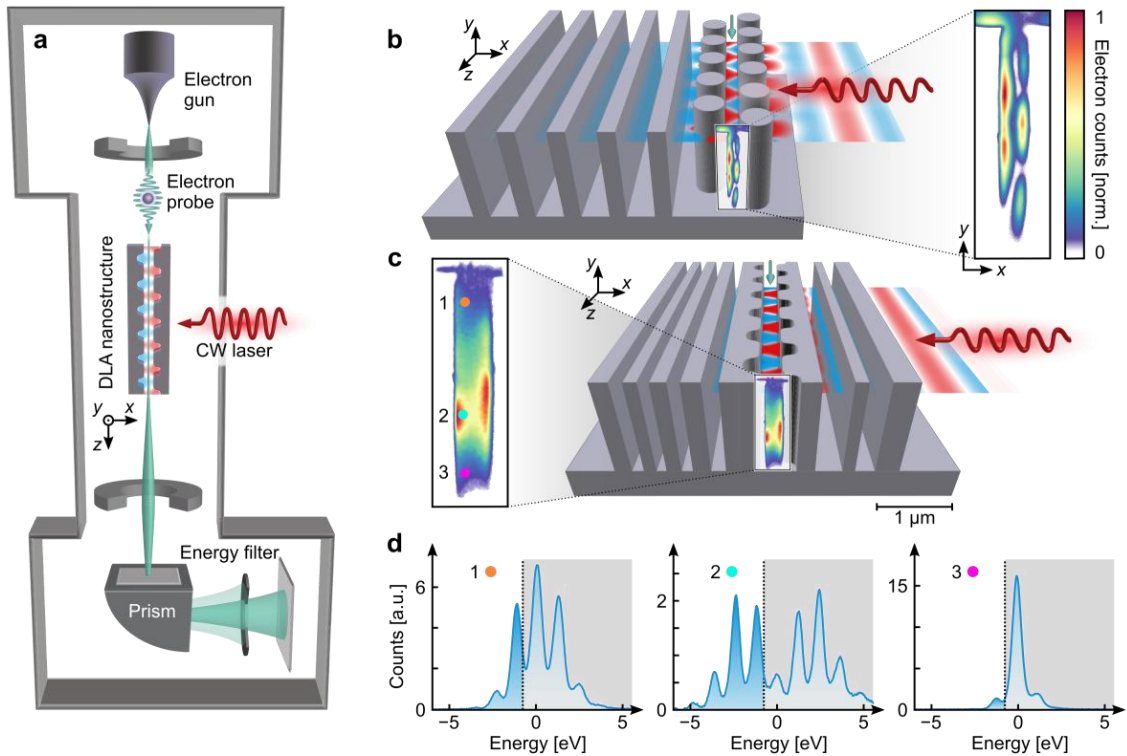

**Figure 1: Imaging the field distribution inside dielectric laser acceleration (DLA) structures. (a)** An illustration of the transmission electron microscope experimental setup. A LaB$_6$ electron source operates in thermionic mode; a set of magnetic lenses and apertures are used to condition the electron beam such that it enters the DLA parallel to the channel at the required spot size; a 1064nm CW laser modulates the energy of the electrons passing through the 20 μm long DLA structure; the electrons continue to a spectrometer, where they are energy-filtered such that only electrons that gained energy are imaged. **(b, c)** 3D model of the two structures used in this work: a dual-pillar structure with a distributed Bragg reflector and an inverse-designed resonant structure, respectively. A representative field distribution is overlaid at the end of the channel. **(d)** Measured electron energy spectra at representative locations inside the channel of the inverse-designed structure. These energy spectra measurements are obtained with an electron beam spot size of ~70 nm, which is smaller than the channel widths of 210 and 280 nm in the two DLA structures. In contrast, the electron beam used to image the field distributions (insets of b and c) has a spot of ~3 μm, which is larger than the dimensions of the channels in all directions. The gray region was filtered out to obtain the images of the acceleration field profile.

The capability we developed for such imaging is inspired by recent advances in the field of photon-induced nearfield electron microscopy (PINEM)[33–36]. To visualize the nearfield of a nanostructure, PINEM relies on free electrons exchanging energy with the nearfield of the nanostructure in units of the light quanta. By collecting only the electrons that gained energy, the nearfield of the nanostructure is imaged with deep-subwavelength resolution. In contrast to our work, most previous PINEM works have relied on the synchronized interaction of electron pulses and lasers pulses[33,35,37,38]. Specifically, the pulsed nature of the interaction has recently been used to visualize a variety of physical phenomena in nm spatial and fs temporal resolutions[10,39–47]. While most works have necessitated a pulsed interaction to reach a strong enough intensity, a few recent works have shown the PINEM-type interaction with continuous-wave (CW) light[31,48–52]. Our work demonstrates imaging by CW-driven PINEM, which enables the unique combination of deep-subwavelength spatial resolution with a sub-nm spectral resolution.

## The experimental setup

Our system (Fig. 1a) is based on a TEM in which thermally emitted electrons are accelerated to an energy of 189 keV and enter the DLA structure (more on the system in Ref. [31]). A 1064 nm CW laser is focused onto the structure. After the interaction with the optical nearfield in the DLA channel, the emerging electrons reach an electron energy loss spectrometer (EELS), which uses a tunable energy filtering slit and magnetic lenses to create an image of those electrons that gained energy. This way, an image of the acceleration profile $|g(x,y)|$ in the channel is generated $|g(x,y)| = \left| \int_{-\infty}^{\infty} E_z(x,y,z) e^{-i\omega z/v_e} dz \right|$ where $E_z(x,y,z)$ is the nearfield distribution of the electric field along the electron trajectory, $\omega$ the angular frequency, and $v_e = \beta c \approx 0.69c$ is the electron velocity. The phase term $e^{-i\omega z/v_e}$ thereby means that the electron samples the specific spatial Fourier mode of the field $E_z^q(x,y,z) = E_z^q(x,y) e^{iqz}$ with $q = \omega/v_e$, which is the mode of interest for electron acceleration.

The Fourier mode $q = \omega/v_e$ imaged by our technique is thus determined by the electron velocity and the laser wavelength, which can be tuned to map the spectral response, as shown below. If the electron velocity fulfills the synchronicity condition $\lambda = \Lambda/\beta$ for the laser wavelength $\lambda = 2\pi c/\omega \approx 1064$ nm and the structure periodicity $\Lambda = 733$ nm, the signal is strongest. The relatively uniform illumination allows us to neglect edge and dephasing effects such that the measured acceleration profile is proportional to the field distribution $\left| E_z^{\omega/v_e}(x,y) \right|$ of a single unit cell of the DLA structure, scaling linearly with the number of periods. We note that the DLA structures are designed to have their field dominated by the Fourier mode that contributes to the acceleration, that is, $E_z(x,y,z) \approx E_z^{\omega/v_e}(x,y) e^{iz\omega/v_e}$. We henceforth shorten the notation by writing the field without the $\omega/v_e$ superscript.

We probe the field in the channel by two techniques: (1) Using a parallel electron beam with a small spot size of ~70 nm, smaller than the channel width, we measure the electron energy spectrum at various selected locations in the channel (Fig. 1d). This approach could be extended to image the field distribution by scanning over the channel (a smaller electron spot size could be used for better resolution, but at the expense of lowering the electron flux). (2) We tune the magnetic lenses controlling the electron divergence and spot size of the TEM to create a collimated electron beam that illuminates the DLA channel homogeneously (with a 3 μm spot size). To acquire an image of the accelerating field, we collected only electrons that gained energy, while filtering out both electrons with no interaction (namely, the zero-loss electrons) and the electrons that lost energy in the interaction. To determine the filtering range (gray region in Fig. 1d), we rely on the first technique to find the optimal energy filter that yields a maximal image contrast.

Scanning electron microscopy (SEM) images of the dual-pillar and inverse-designed structure are shown in Fig. 2. The electrons (marked in blue) enter the structures via an alignment aperture and are modulated in the structure by the optical field of a CW laser impinging on the structure perpendicularly to the electron flow. The laser illumination area, marked as a red surface, has a Gaussian distribution with spot diameter of ~15 μm ($1/e^2$), which is smaller than the structure length of 20 μm. The insets in Figs. 2a and b show the simulated xz-field distribution $E_z(x,z)$ of an ideal structure. We note that along the x-direction, the field distributions are always designed to have significant non-vanishing values at the center of the channel, where electrons can be guided for acceleration. However, our experimental data reveals that the field inside some of the fabricated structures is far from optimal and can even vanish at the center of the channel.

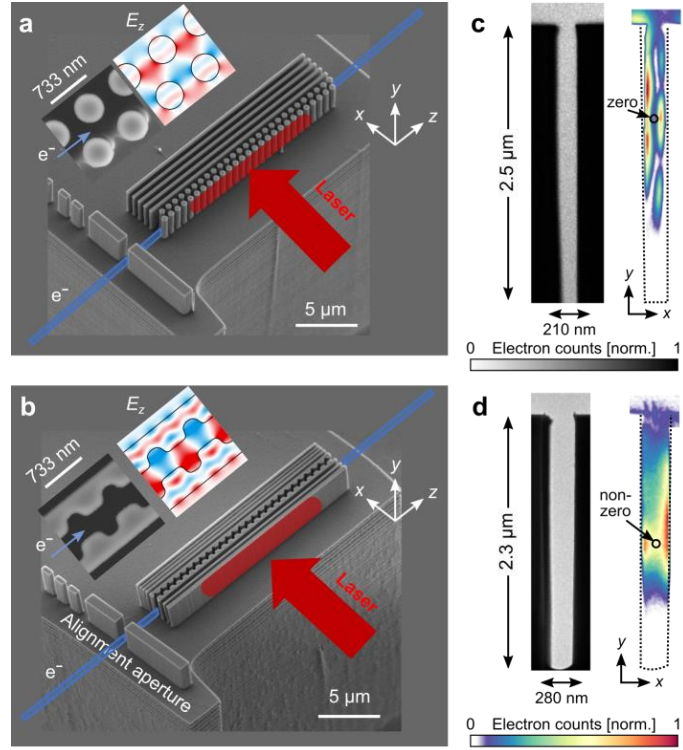

**Figure 2: Scanning electron microscopy images of the dielectric laser acceleration (DLA) structures.** The electrons (blue) enter the structures via an alignment aperture and are accelerated in the structure by the electric field of a 1064 nm CW laser impinging on the structure perpendicular to the electron flow. The laser illumination area is marked as a red surface showing that the spot diameter of 15 μm ($1/e^2$) is smaller than the structure length of 20 μm. **(a, b)** The dual-pillar structure with Bragg mirror and the resonance-enhanced inverse-designed structure, respectively. A zoom-in view into the structure and the corresponding accelerating electric field distribution $E_z(x,z)$ are shown as insets for both structures. The periodicity of 733 nm provides phase matching between the electrons and the electric field along the z-direction. Note that $E_z$ has a significant non-vanishing value at the center of the channel, where electrons are guided. **(c, d)** Typical TEM images of the two DLA channels were taken with a broad 3 μm spot-size electron beam, without (left, black-white color scale) and with (right, rainbow color scale) electron energy filtering. The unfiltered image shows the uniformly illuminated electron channel, while the filtered image shows the PINEM field distribution. The measured field distributions extend above the structure's top and vanish far above the channel's bottom in both cases. In addition, the field measured in the dual-pillar (c) structure has a null at the center, which is different from its design (as discussed below).

## Experimental results and analysis

Figures 2c and d present typical PINEM images of the two DLA channels that were taken without (left) and with (right) electron energy filtering. The unfiltered images show the uniformly illuminated channels, while the filtered images show the PINEM field distribution inside each DLA channel, which is related to $|E_z(x,y)|$ as shown by our theory. The dual-pillar field distribution (Fig. 2c) exhibits several outstanding features. Along the vertical y-direction, the field distribution shows an oscillatory behavior and extends beyond the top of the structure. Furthermore, the field does not extend all the way down to the bottom of the channel. As discussed later, this behavior can be predicted by accurate 3D numerical simulations and can be attributed to the eigenmode profile of the nanostructure sitting on a silicon substrate. For comparison, inspecting the field distribution of the inverse-designed structure (Fig. 2d) reveals intriguing differences and similarities to the dual-pillar structure. Along the vertical y-direction, the field is again not reaching down to the channel bottom, but in contrast to the dual-pillar structure, it does not extend above the channel top, indicating that the field is more strongly confined to the channel. Additionally, the field along the y-direction does not show the oscillatory behavior of the dual-pillar structure.

Along the horizontal x-direction, the measured field of the dual-pillar structure exhibits a null at the channel center (Fig. 2c) – the opposite of the ideal DLA performance. Following Plettner *et al.*[53], we associate this field distribution with an anti-symmetric mode that follows a hyperbolic sine (sinh) profile. As seen later, the dominance of this mode over the designed symmetric mode that follows a hyperbolic cosine (cosh) distribution can be explained by a fabrication mismatch. For comparison, the field distribution of the inverse-designed structure (Fig. 2d) along the horizontal x-direction, at center height, resembles the expected cosh-like symmetric mode distribution with a non-zero value at the center. A dominant symmetric mode is not only required for optimal acceleration, but when looking forward toward practical applications of DLAs, it can also be used for transverse confinement of the electron beam[28,29].

The field distribution was measured over a range of wavelengths (Figs. 3a and b), from 1063 nm to 1065.4 nm, controlled by changing the CW laser operating temperature (Supplementary Fig. S1). Interestingly, the field distribution of the dual-pillar structure hardly changes with wavelength (Fig. 3a). However, for the inverse-designed structure, the field distribution changes dramatically, with its peak moving from one side to the other.

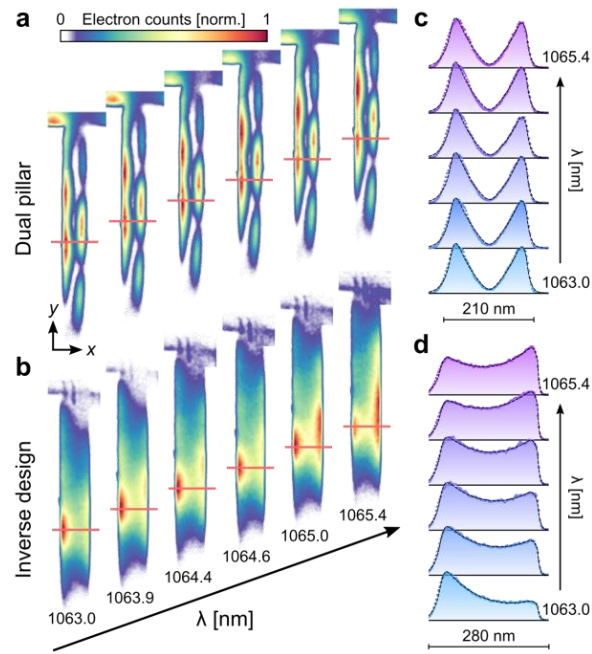

**Figure 3: Measured spectral response of the field distribution inside the DLA channel.** Our method for deep-subwavelength spatial imaging with sub-nm spectral resolution is applied to **(a)** the dual-pillar structure and **(b)** the inverse-designed structure. We observe a significant change for the inverse-designed structure and hardly any change for the dual-pillar structure. **(c, d)**

Measured field cross section (blue/purple dots) along the red marking lines in (a) and (b). The black dashed lines (which can hardly be distinguished from the blue/purple dots since they almost perfectly overlap) are fits of the expected cosh/sinh field profile within the channel. They were multiplied by an exponential decay at the edges of the channel, where electrons are attenuated by inelastic scattering off the structure. A Gaussian convolution imitates the blurring effect mainly due to the imaging point spread function. The dual-pillar structure has a large sinh component in the fit, which corresponds to an anti-symmetric (odd) structural mode with zero field at the center. In contrast, the inverse-designed structure has a significant cosh component in the fit, with large non-vanishing energy at the channel center corresponding to a low-order symmetric (even) mode.

Figures 3c and d show that both measured symmetric and anti-symmetric field profiles match with theory[53] and can be approximated accurately with a combination of cosh and sinh functions. When comparing the field distribution to a PINEM image, one further has to consider blurring effects that arise mainly due to the imaging point spread function that we model by a Gaussian convolution. Combining the Gaussian spread with the sinh and cosh mode profiles takes the form $\exp(x^2/2\sigma^2) * [A\cosh(\kappa x) + B\sinh(\kappa x)]$. The coefficients $A$ and $B$ and the evanescent decay rate $\kappa$ are determined by a fit, while the root-mean-square of the Gaussian was chosen as $\sigma = 10$ nm for optimal fit results. At the edges of the channel, electrons lose energy due to inelastic scattering off the silicon boundaries. This scattering results in a drop of signal at the boundaries in our PINEM images, which we model by a fitted exponential decay.

Our model (filled dashed curves in Fig. 3c and d) yields excellent agreement with the experimental data (blue/purple dots in Fig. 3c and d) over the entire spectral range that was measured. The measured results show that the field distribution in the dual-pillar structure corresponds almost purely to the anti-symmetric mode with a large sinh contribution in the fit ($A/B \approx 1/10$), whereas the field in the inverse-designed structure corresponds to the symmetric mode distribution with a dominant cosh component in the fit ($A/B \approx 50$). As seen next, the deviation of the dual-pillar field from the design can be explained by a change in the actual pillar diameter relative to the intended design.

To understand the observed field profiles, we performed extensive 3D numerical simulations to study the sensitivity of the field distribution to deviations in the structure dimensions. Figures 4a and b show the field distributions inside the two structures for small variations of the structure diameters. We singled out the structure diameter $D$ (left column in Fig. 4) as the parameter which best describes the fabrication inaccuracy originating from a too long or too short e-beam illumination time during the lithography process. The best match with the experimental data is obtained for $\delta D = -48$ nm (where $\delta D$ is the change in $D$) for the dual-pillar structure (that is, a 48 nm narrower pillar diameter than the designed diameter of 457 nm) and $\delta D = 0$ nm for the inverse-designed structure.

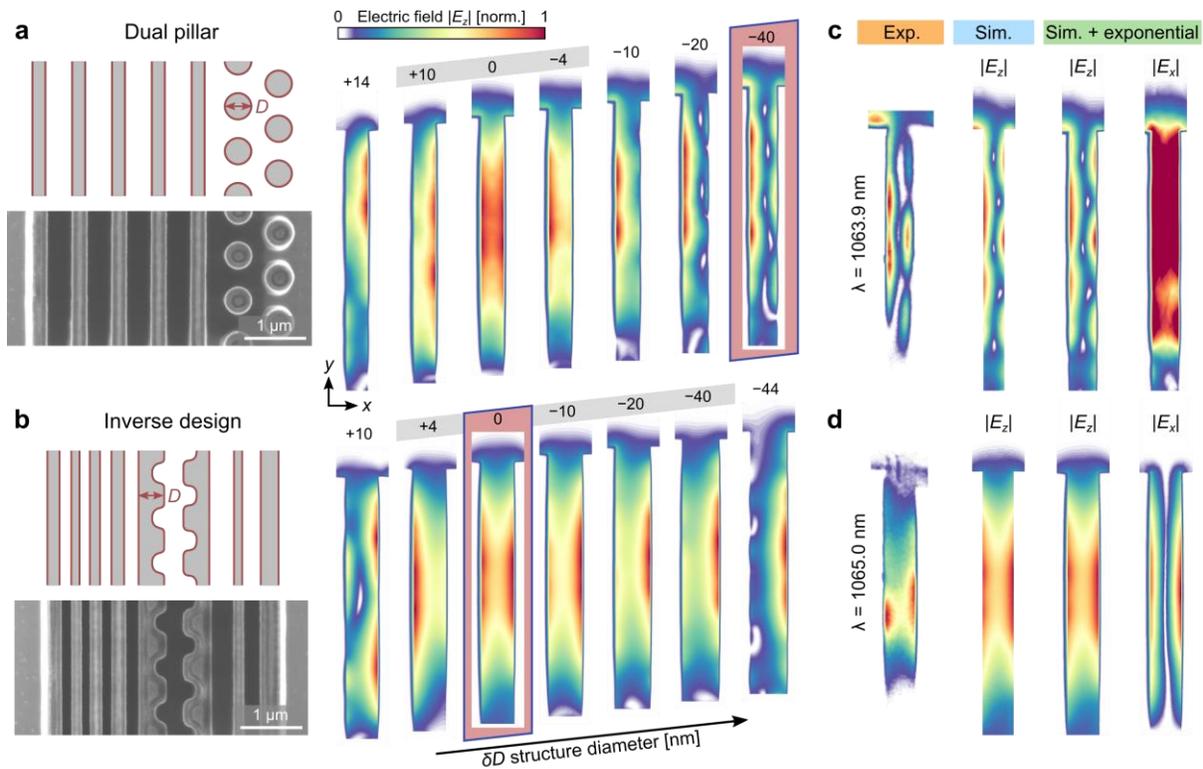

**Figure 4: Sensitivity of the DLA nearfield to changes in structure geometry.** The accelerating field distribution $|E_z(x,y)|$ is investigated by 3D simulations as a function of deviations $\delta D$ from the nominal size $D$, marked in the SEM images on the left. **(a, b)** Dual-pillar and inverse-designed structures, showing field simulations for a scan over $\delta D = +14$ to $-40$ nm and $\delta D = +10$ to $-44$ nm, respectively. The best fit results are obtained for $\delta D = -48$ nm and $\delta D = 0$ for the dual-pillar and inverse-designed structures, respectively. Near the design target ($\delta D = 0$), the symmetric mode is dominant for both structures, as designed. For small changes $\delta D$, the field maxima move from side to side with increasing contribution from the anti-symmetric mode. Close to the best fit result of the dual-pillar structure ($\delta D = -40$ nm, marked by red rectangle), its anti-symmetric mode is highly robust against changes in $\delta D$. These simulation results match the experimental findings of Fig. 3, where the symmetric mode of the inverse-designed structure moves from side to side, whereas the anti-symmetric mode of the dual-pillar structure does not change much with wavelength. **(c, d)** Side-by-side comparison of measured and best-fit simulated fields. Looking at $|E_x(x,y)|$ reveals that a symmetric mode in $E_z$ is associated with an anti-symmetric mode in $E_x$ and vice versa.

While both structures show the symmetric mode around the design target ($\delta D = 0 \pm 5$ nm), this region is highly sensitive to the structure diameter, and the peak acceleration is seen to shift to one side as the contribution from the anti-symmetric mode increases. This matches our observation from Fig. 3b, where the mode profile of the inverse-designed structure shifted with wavelength. In contrast, our simulations of the dual-pillar structure indicate that the anti-symmetric mode found around $\delta D = -48$ nm is robust to changes in $\delta D$. This robustness agrees with our observation from Fig. 3a, displaying no sensitivity of the dual-pillar field profile to changes in wavelength. Altogether, the inverse-designed structure is far more robust to changes in the structure dimensions because its symmetric mode stretches over a much wider range ($+4$ nm $> \delta D > -40$ nm) than that of the dual-pillar design ($+10$ nm $> \delta D > -4$ nm). Our two fabricated structures are a testimonial of the robustness of the inverse-design approach.

Figures 4c and d demonstrate the amount of knowledge that can be gained when combing our experimental nearfield measurement with 3D simulations. The measured profile of the accelerating field $|E_z|$ is shown next to the simulated field for the structure parameters that give an optimal replication of the measurement. We added an exponential decay at the vacuum-silicon interface to account for the inelastic scattering of the electrons off the boundaries of the channel. Having found the matching structure parameters this way, we can then extract from the simulation the other field components, $|E_x|$ (rightmost picture) and $|E_y|$ (see Supplementary Fig. S3). This represents a major advantage over the traditional approach of predicting the field distribution based on measured structure dimensions. Measurements by SEM, for example, can mainly identify the dimensions of outer features of the structure but often misses information inside the nanostructure, since it requires to manually cut the structure at the specific imaging location which is a priory unknown.

The simulations of $E_z$ in Figs. 4c and d capture a wide range of details observed in the measurement. Along the horizontal x-direction, we retrieve the anti-symmetric profile for the dual-pillar structure and the symmetric profile for the inverse-designed structure. Along the vertical y-direction, the simulation of the dual-pillar structure reproduced the oscillating pattern as well as the extension of the accelerating field above the left row of pillars. As seen experimentally, the inverse-designed structure does not exhibit this feature above the structure. Lastly, the simulations also reflect the observation of the field not reaching the channel bottom. The slight deviation seen in the field strength at the bottom of the channel can be ascribed to a small reduction in electron counts coming from the Gaussian beam shape of width 3 µm. These results manifest the need for 3D modeling of these devices, instead of a 2D model that usually approximates the structure as invariant along the vertical direction for convenience and ease of calculations.

The match between experiment and theory allows us to use the simulation for the analysis of the deflecting field $E_x$ (Figs. 4c and d). We found that a symmetric mode in $E_z$ is accompanied by an anti-symmetric mode in $E_x$, and vice versa, which is in accordance with the existing theory[53]. $E_y$ is close to zero throughout most of the channel and only has significant contribution above the structure (see Supplementary Fig. S3). Thus, not only does the anti-symmetric mode in $E_z$ lead to zero net acceleration, it also strongly deflects the electrons sideways. In contrast, the symmetric mode in $E_z$ has a sinh mode distribution in $E_x$ that can be used to focus the beam along the x-direction, or defocus it, depending on the delay between the electron arrival time and the phase of the field[28,29]. A possible application of the anti-symmetric $E_z$ modes in future DLA structures could be for beam deflection.

We have so far seen that many of the observed features in the nearfield distribution can be attributed to the 3D nature of the DLA structure. It is therefore interesting to compare the simulated 2D and 3D results. Supplementary Fig. S2 shows that in the 2D simulation, the inverse-designed structure is expected to have a narrower and stronger spectral response, which can be explained by its resonant enhancement. Notably, the peak acceleration gradient of the inverse-designed structure is higher despite its channel width of 280 nm being larger than that of the dual-pillar structure of 210 nm. In the 3D simulation, the inverse-designed Fabry–Perot-like resonator, which has a height comparable to the wavelength, suffers higher scattering loses, resulting in a drop in quality factor. This is reflected in a broader spectral response and lower acceleration gradient. These findings highlight the need for 3D simulations to correctly capture the spectral response.

## Discussion and outlook

We developed a technique based on continuous-wave PINEM to image the field distribution inside the channels of the two leading DLA structures, achieving deep-subwavelength spatial resolution and showing their spectral response with sub-nm wavelength resolution. While the inverse-designed structure agreed well with its designed symmetric (cosh) field distribution along the horizontal direction, the dual-pillar structure strongly deviated from it and showed a dominant anti-symmetric (sinh) field distribution. By a 3D numerical analysis, we replicated the measured field distributions along both the horizontal and vertical directions and retrieved all three field components in real space. We were able to attribute the measured deviations to a specific fabrication mismatch of the structure diameter (being 48 nm thinner than designed), showing the precision and prospects of our experimental technique. Our investigation highlighted the robustness of the inverse-designed structure to changes in the structure dimensions, which is a key advantage in future DLA experiments.

For complex DLA designs with varying periodicity along the electron propagation[18,28,29], we envision extending our approach to a full 3D tomography of the field inside the device. To this end, we propose to characterize the field at individual cross-sections along the device by selectively illuminating one sub-section at a time. Only the illuminated sub-section will take part in the acceleration process, and we can extract the local nearfield in that sub-section from the measurement. One way to achieve this could be integrating apertures onto the silicon structures. These test structures would sit on the same silicon wafer as the actual DLA structure but would only be used for 3D field characterization. An alternative method is to use a focused laser beam that is scanned along the structure. Both approaches are schematically illustrated in Fig. 5 and simulated in Supplementary Fig. S5. For further study of the nearfield, one can also perform dark-field imaging to measure the transverse field $E_x$.

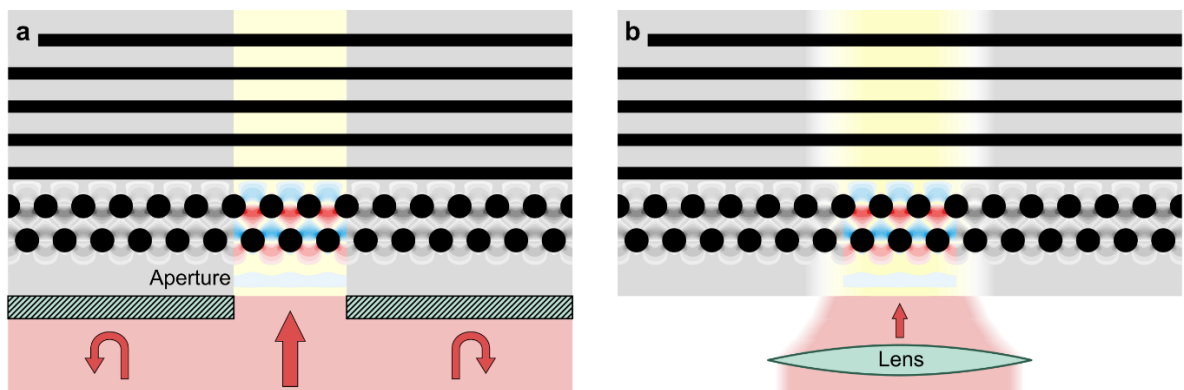

**Figure 5: Proposal for 3D tomography of the field distribution inside a DLA structure.** To gain information about how the nearfield behaves along the length of the structure, we suggest illuminating sequentially individual sub-sections, one at a time, to reconstruct the full 3D field. This can be either achieved by **(a)** including an aperture on the chip design. For each position, a dedicated test structure would be added to the chip. **(b)** Alternatively, a lens can be used to focus the laser to a small spot of a few μm. Scanning the beam along the structure would then provide full 3D field information. A deconvolution algorithm can further increase the spatial resolution along the channel. Simulations of the two approaches are shown in Supplementary Fig. S5.

Our experimental results and their match with simulations shed new light on the predictions about the pros and cons of the two most famous types of DLA structures. While the resonant behavior of the inverse-designed structure originally suggested a stronger acceleration over a narrower bandwidth, we found that it provides a broader bandwidth over a wider range of wavelengths and a wider range of fabrication tolerances. The maximum acceleration gradient at the design parameters is comparable to that of the dual-pillar structure, despite the much broader channel of the inverse-designed structure. For further improvement, a full-3D inverse-design optimization is needed[27].

Looking ahead, the powerful insight gained by continuous-wave PINEM will be of great benefit in the optimization of future DLA designs. More generally, our method can be applied for deep-subwavelength imaging of the field distribution inside many other complex nanostructures and microstructures based on various electron-photon interactions, such as electron wave function shaping structures[54]. Furthermore, one can further perform pulsed-laser and pulsed-electron excitations to investigate the dynamic response of such structures.

# Methods

Experimental setup and fabrication

The experiments were performed in a JEOL JEM-2100 Plus TEM equipped with a Gatan GIF system and the LaB$_6$ electron filament in thermal emission mode. The TEM was modified to couple light into the sample (Fig. 1) [10,45,47].

Data acquisition is done in the converged beam electron diffraction (CBED) mode at low magnification. The electron beam diameter in the focal plane was 3 µm for the PINEM measurements and 70 nm for the point-by-point spectrum analysis with a convergence angle of 0.3 mrad and a 0.6 eV FWHM zero-loss energy width. We aligned the nanostructure channel to the electron trajectory using a double-tilt holder (Mel-Build Hata Holder), whose position relative to the electron beam and the tilt angle is determined with nanometer resolution and a step of 0.1°, respectively, using a custom cartridge to avoid shadowing the optical beam. The energy-filtered TEM (EFTEM) image was obtained by using a 7 eV wide mechanical filtering slit that is located at the exit plan of the EELS prism (Fig. 1). The slit acts as a band-pass filter letting only electrons through that gained energy in the range of -1 to -8 eV, which are then imaged on the Gatan US1000 camera using a pixel binning of x4. The integration time was ∼ 60s .

A 100 mW CW-driven distributed feedback (DFB) laser (QLD106p-64D0) emitting nominally at 1060 nm served as a seed for a two-stage Yb fiber amplifier with a 4 nm filter between the two stages. The laser beam was focused with a cylindrical lens to achieve a 15 µm Gaussian spot ($1/e^2$) and attenuated to 500 µW impinging perpendicular to the electron flow. To tune the laser wavelength, the temperature of the built-in TEC was swept. Figure S1 shows the measured beam spectral distribution for different TEC temperatures.

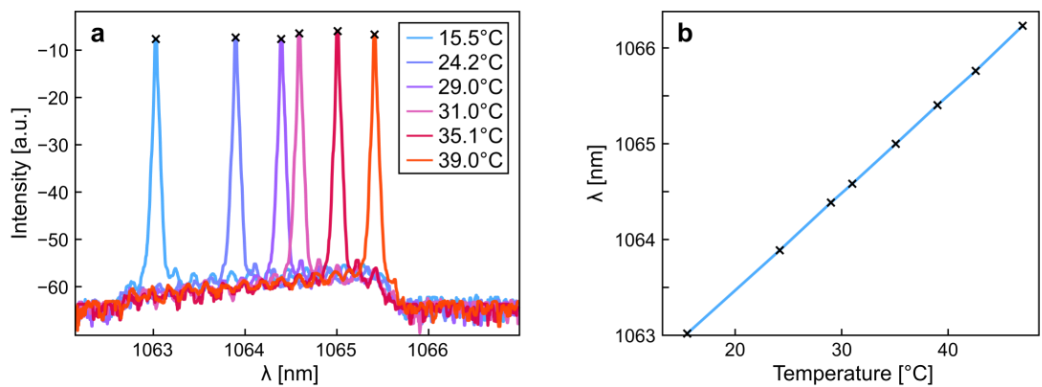

**Figure S1: Measured laser spectrum vs. thermo-electric cooler temperature. (a)** Spectra at various temperatures. **(b)** Peak wavelength vs. temperature extracted from (a). The blue solid line is a fit that confirms the linear scaling.

The 2.5±0.1 µm high nanostructures were fabricated by electron beam lithography (100 kV) and cryogenic reactive-ion etching of 1-5 Ωcm phosphorus-doped silicon[26]. To provide clearance for the laser beam and the electron beam, a 30-µm-high mesa was formed by etching the surrounding substrate away.

Photonic inverse design and 3D simulations

We used an open-source Python package based on a 2D-FDFD simulation (Fig. S2a)[55]. The structure was optimized over a 5-μm-wide design region in the xz-plane, containing a 250-nm-wide vacuum channel in the center for the electrons to propagate through. Periodic boundaries were applied along the z-direction, enforcing the periodicity of 733 nm, and perfectly matched layers were defined along x. A transverse-magnetic plane wave was excited from one side, and the resulting electric field $E_z$ was computed in the center of the channel to find the acceleration gradient, which served as the objective function of the optimization [27,56].

3D-FDFD simulations were performed in COMSOL. The simulation cell (Fig. S2b) consists of one unit cell of the structure with periodic boundaries along the z-direction. The 2.5-μm-high structure sits on a mesa, with the electron channel 5.5 μm away from the mesa edge. Perfectly matched layers enclose the simulation cell along the x- and y-direction. A Gaussian beam with a beam waist of 15 μm FWHM is focused onto the center of the channel at a height of 1.8 μm above the mesa. An angle of incidence of 5° above the substrate was observed to reproduce the experimental data best and was therefore chosen.

Figure S2c shows the 2D simulated electric field $E_z$ in the xz-plane for the two structures. Figures S2d and e show the spectral response of the dual-pillar (dashed line) and the inverse-designed (solid line) structures from 2D and 3D simulations, respectively. The 2D simulated spectrum reveals the maximum acceleration gradient at the designed wavelength of 1064 nm. The bandwidth of the inverse-designed structure is twice as narrow as that of the dual-pillar structure, indicating the resonant nature of the former. By contrast, the 3D simulation shows a slightly narrower response of the dual-pillar structure, which is in accordance with the observed narrower cosh-like mode with respect to changes in structure diameter from Fig. 4. The stark difference between the 2D and 3D results are a testimonial to the importance of 3D simulations, and the excellent agreement of our experimental data with the 3D results confirms their accuracy.

Another difference is that the 3D resonance frequencies of the inverse-designed and dual-pillar structures are shifted to 287 THz (1045 nm) and 284 THz (1056 nm). Ultimately, a 3D inverse-design optimization should be performed.

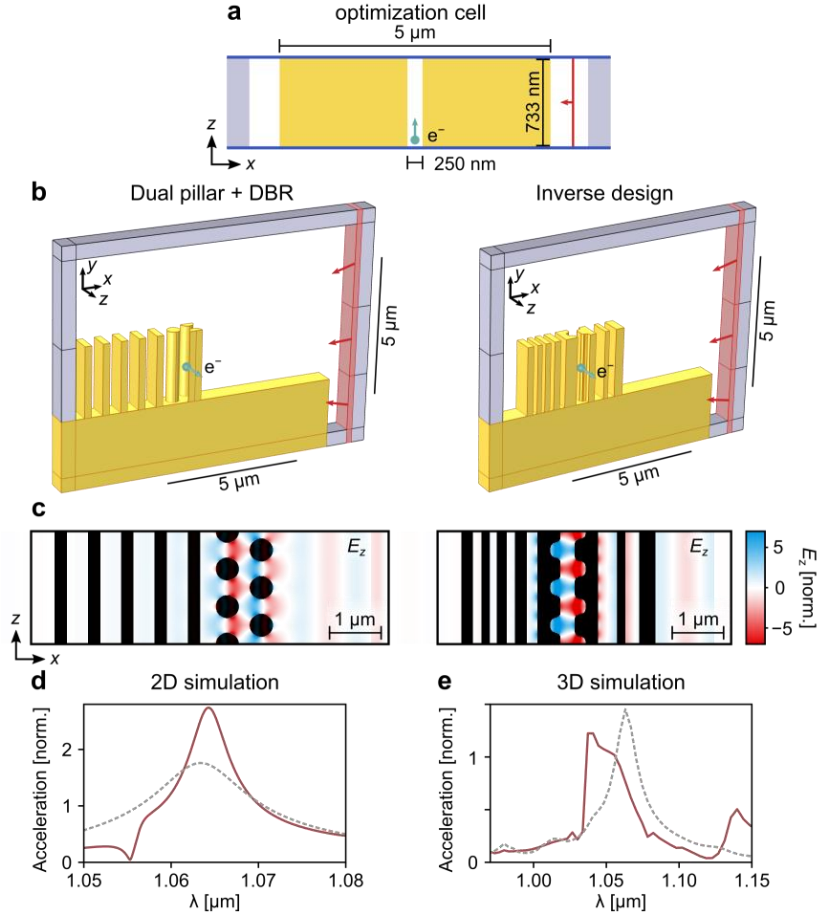

**Figure S2. Comparison between 2D and 3D simulations. (a)** Two-dimensional simulation cell as used for the inverse-design optimization and the spectral response (d). A single unit cell of the grating is optimized. Periodic boundaries (blue) along the longitudinal direction are applied to simulate an infinitely long grating with a periodicity of 733 nm. The design region (yellow) is 5 µm wide and includes a 250-nm-broad vacuum channel (white) in the middle. A plane wave (red) is impinging from the right-hand side. Perfectly matched layers (gray) are used to absorb light and imitate free space. **(b)** The simulation cells for the 3D-FDFD simulations of the dual-pillar (left) and inverse-designed (right) structures. The 2.5-µm-tall nanostructure sits on a mesa and a Gaussian beam is focused onto the center of the electron channel. **(c)** Electric field distribution $E_z(x, z)$ for the dual-pillar (left) and inverse-designed (right) structures, obtained by 2D simulation. **(d, e)** Calculated spectral response of the dual-pillar (dashed, gray line) and inverse-designed (solid, red line) structure from 2D and 3D simulation, respectively. While the 2D simulation (d) predicts a stronger acceleration gradient and a narrower spectral response for the inverse-designed structure, the two structures perform similarly in 3D (e).

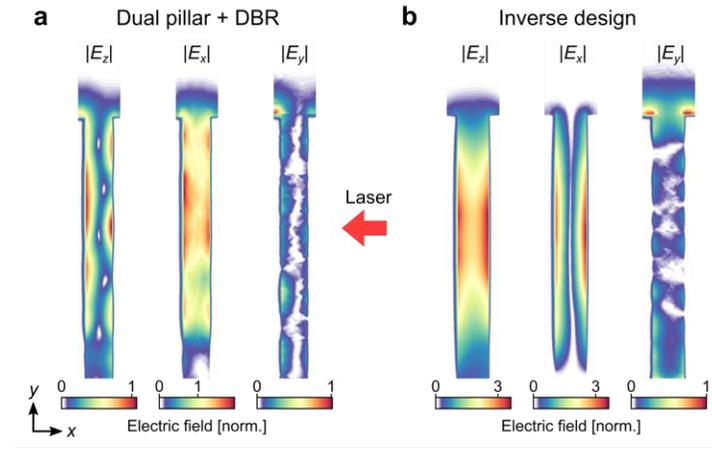

**Figure S3. Simulated field components inside the DLA channel.** The field simulations for the design parameters which best reproduce the experimental field measurement is shown for the **(a)** dual-pillar and **(b)** inverse-designed structure. It demonstrates that an anti-symmetric mode in $E_z$ is accompanied by a symmetric mode in $E_x$ and vice versa. The electric field $E_y$ along the vertical direction is close to zero inside the channel and only strong above the structure.

Data post-processing

The following steps were applied for post-processing of the PINEM images (see Fig. S4):

1) Rotation: The original picture was rotated by quadratic interpolation such that the channel is vertically oriented.
2) Dark line removal: In our Gatan imaging filter, the mechanical filtering slit had a local contamination that created a dark line parallel to the horizontal axis. We removed the line by quadratic interpolation of nearby points along the vertical direction. This step was not necessary for the unfiltered data, for which the slit was fully opened.
3) Cropping: The picture was cropped to the region of interest.
4) Colormap change to highlight the relevant non-zero field distributions.

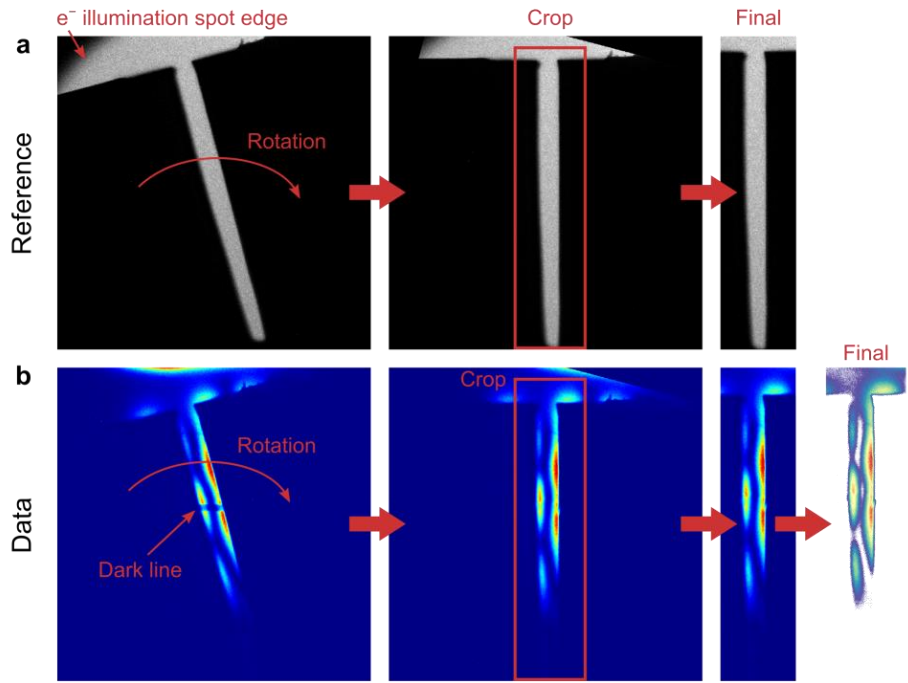

**Figure S4. Process of data manipulation. (a)** Unfiltered image as reference. **(b)** Energy-filtered TEM (EFTEM) image. First, the images are rotated by a bilinear transformation. Second, a dark line artifact in the filtered image is removed by linear extrapolation of the neighboring points.

Proposed tomography method

To image the field distribution in subsections of the structure, we suggest to either place an aperture in front of the area of interest or to focus the laser beam onto that region. Figure S5 shows 2D simulations for those two scenarios and confirms that the acceleration profile closely follows the excitation beam shape. As long as the beam is not focused too tightly or the aperture is too small, diffraction effects are minor. In the proposed 3D tomography method, the laser beam is scanned along the structure length, or the aperture is placed at multiple locations along the structure. A deconvolution algorithm is then applied to retrieve the field information as a continuous function of z. For the acceleration of sub-relativistic electrons, the structure periodicity is tuned to maintain synchronicity as the electron velocity increases. To image the field at a farther point along the structure, it is therefore necessary to pre-accelerate the electrons to the velocity of each subsection.

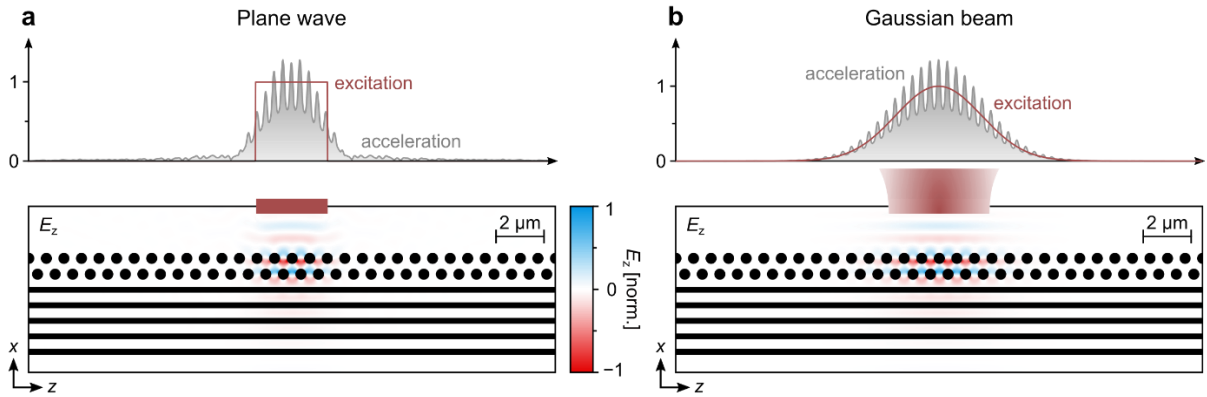

**Figure S5: Proposed illumination schemes for 3D field tomography.** The acceleration profile is shown together with the shape of the excitation laser beam. Underneath is the 2D field simulation in the xz-plane. The two proposed schemes are **(a)** a plane-wave excitation as achieved by placing an aperture in front of the structure, and **(b)** a focused Gaussian beam. In both cases, the acceleration profile closely follows the excitation.